\documentclass[preprint]{aastex}
\pdfoutput=1

\newcommand{\wcen}{$\omega$~Cen}
\newcommand{\kms}{km s$^{-1}$}

\slugcomment{Accepted for publication in the Astronomical Journal}

\shorttitle{Radial Velocity Survey for $\omega$ Centauri Members}

\shortauthors{Da Costa and Coleman}

\begin{document}

\title{A Spectroscopic Survey for $\omega$ Centauri Members at and beyond the Cluster
Tidal Radius}

\author{G. S. Da Costa\altaffilmark{1}}
\altaffiltext{1}{Research School of Astronomy \& Astrophysics, The Australian National 
University, Mt~Stromlo Observatory, via Cotter Rd, Weston, ACT 2611,
Australia; gdc@mso.anu.edu.au}
\and
\author{Matthew G. Coleman\altaffilmark{1,2}}
\altaffiltext{2}{Max-Planck-Institut f\"{u}r Astronomie, K\"{o}nigstuhl 17, D-69117 Heidelberg, 
Germany; coleman@mpia-hd.mpg.de}

\begin{abstract}
We have used the two-degree field (2dF) multi-fibre spectrograph of the Anglo-Australian Telescope
to search for candidate members of the unusual globular cluster $\omega$ Centauri at and beyond
the cluster tidal radius.  Velocities with an accuracy of $\sim$10 \kms\/ were obtained for
4105 stars selected to lie in the vicinity of the lower giant branch in the cluster color-magnitude
diagram and which cover an area on the sky of $\sim$2.4 $\times$ 3.9 deg$^2$ centered on the
cluster.   Within the velocity interval 190--270 \kms, 
the cluster member candidates have a steeply declining surface density distribution consistent with 
the adopted cluster tidal radius of 57$\arcmin$.
For the stars in the sample beyond the tidal radius, an analysis of line-strengths from the spectra, as well as radial velocities, identifies only {\it six}\/ stars as possible candidates for extra-tidal association with 
the cluster.
If all six of these stars are indeed related to the cluster, then a maximum of 0.7 $\pm$ 0.2 per
cent of the total cluster mass is contained in the region between one and two tidal radii.  
Given this limit, we conclude that there is no compelling evidence for any significant 
extra-tidal population in \wcen.  The effects of tidal shocks on the outer parts of the
cluster are consistent with this limit.
Theories for the origin of \wcen\/ frequently suggest that the
cluster is the former nucleus of a tidally stripped dwarf galaxy.  Our results require 
that the stripping process must have been largely complete at early epochs, consistent with
current dynamical models of the process.  The 
stripped former dwarf galaxy stars are therefore now widely distributed around the Galaxy
\end{abstract}


\section{Introduction}

The stellar system $\omega$ Centauri is unusual in many respects.  When considered as part of
the Galaxy's globular cluster population it is among the most luminous and the most massive,
with estimates for the cluster mass ranging from 2 to 5 $\times$ 10$^{6}$ M$_{\sun}$ 
\citep[e.g.][]{vdv06, MM95}.  
Unlike other globular clusters though, \wcen\/ exhibits the characteristics of system that has
undergone substantial chemical self-enrichment, with its member stars showing a wide spread
in abundance \citep[e.g.][]{FR75, ND95, EP02},  and complex distributions of 
element-to-iron abundance ratios \citep[see][and the references therein]{DR07}.  The 
abundance picture is further complicated by the discovery that the lower main sequence 
of the cluster is bimodal \citep{BD04}, with the bluer sequence having a higher abundance 
\citep{Pi05}.  These observations can be explained if the metal-rich but bluer sequence has a substantially higher helium abundance \citep[Y $\approx$ 0.40, see:][]{JN04, Pi05}, but generating such 
large helium abundances within the context of chemical evolution models is difficult \citep[e.g.][]{DR07}.
An extensive spectroscopic survey of \wcen\/ post-main-sequence members that characterizes
the many and diverse properties of this cluster is presented in \citet{vL07}. 

The dynamics of \wcen\/ are also unusual.  Unlike most globular clusters, it has a notably flattened 
shape and rotates comparatively rapidly \citep[see][and the references therein]{vdv06}.  Moreover, 
the stellar abundances and stellar kinematics are correlated, with the more metal-rich stars
being not only more centrally concentrated but also kinematically cooler \citep[e.g.][]{NF97, SP05}, 
although see also \citet{Pa07}.  Possible substructures with the cluster may also exist.
For example, \citet{vdv06} find evidence in their kinematic analysis for a separate disk-like component between 1 and 3 arcmin from the cluster center.  Finally, the orbit of \wcen\/ is also unusual in that it is
tightly bound and {\it retrograde}\/ with apo- and peri-galactocentric distances of 6.2 and 1.2 kpc, 
respectively, and a period of 120 Myr \citep{D99b}.  The cluster never rises very far above the Galactic 
plane: \citet{D99b} give $z_{max}$ as $1.0 \pm 0.4$ kpc.

These characteristics have led to increasing support for the idea \citep[cf.][]{F93,NFM96,NF97} 
that \wcen\/ has not evolved in isolation, but in fact represents the nuclear remnant of a now disrupted nucleated dwarf galaxy that has been accreted by the Milky Way.  Models of the orbital
decay and disruption of a nucleated dwarf, which is presumed to lie initially far from the Galaxy's 
centre, have shown 
that this process is dynamically plausible, with most of the dwarf's mass lost as the 
nucleus settles into an orbit like that of the present-day \wcen\/ \citep{BK03, TD03}.  
These dynamical models, however, do not cover a full Hubble time and while the likely present-day
outcome is that the surviving nucleus  is an isolated bound remnant, one may seek to confirm this 
prediction observationally.  For example, in the model
of \citet{BK03}, the mass of the stellar envelope surrounding the nucleus, which comes from
the disrupting dwarf, is still fully $\sim$50\% of the mass of the nucleus at the end of the calculations
2.6 Gyr after the initiation of the interaction.  At this point the nucleus has achieved an orbit comparable
to that of \wcen.  The most likely signature of any remnant from such an interaction would be in the 
form of structure beyond the tidal radius of the cluster (nucleus), probably in the form of tidal tails.

Extra-tidal structure, particularly in the form of tidal tails, is well established around a number of 
globular
clusters \citep[e.g.][and the references therein]{G95,LMC00,GJ06}, with the most striking example 
being Pal~5 \citep{MO03, GD06}.  The origin of these features though lies with the interaction between
the cluster and the tidal field of the Galaxy rather than as a remnant from any merger process.
Tidal tails have been claimed to exist in \wcen\/ by \citet{LMC00} who analyzed the spatial
distribution of color-magnitude diagram selected stellar samples, chosen to maximize the contrast
between field and cluster stars in their photographic data.  For \wcen\/ they 
found ``two large and significant tidal tails'' extending up to $\sim$2 degrees to the north and south 
of the cluster and estimated that the features contained up to 1\% of the cluster mass.  They noted,
however, that dust absorption, which they did not account for, might affect the star counts.
Subsequent work by \citet{Law03} \citep[see also][]{MC07}, however, showed that these features are
much less prominent when reddening variations are properly accounted for.

In this paper we present the results of an extensive spectroscopic survey in which we use radial
velocity and line strength information to investigate the number and distribution of \wcen\/ stars
both within and beyond the cluster tidal radius, and to place constraints on the outer structure of the
cluster and on any extra-tidal population present.
The following section describes the selection of the sample, the observations and the derivation of the
radial velocities.  Section 3 provides the analysis for stars both in the cluster and outside
the cluster tidal radius.  The main results are discussed in \S 4 and the conclusions are summarized in
\S 5.

\section{Observations and Reductions}

\subsection{Sample Selection}

\begin{figure}
\epsscale{1.0}
\plotone{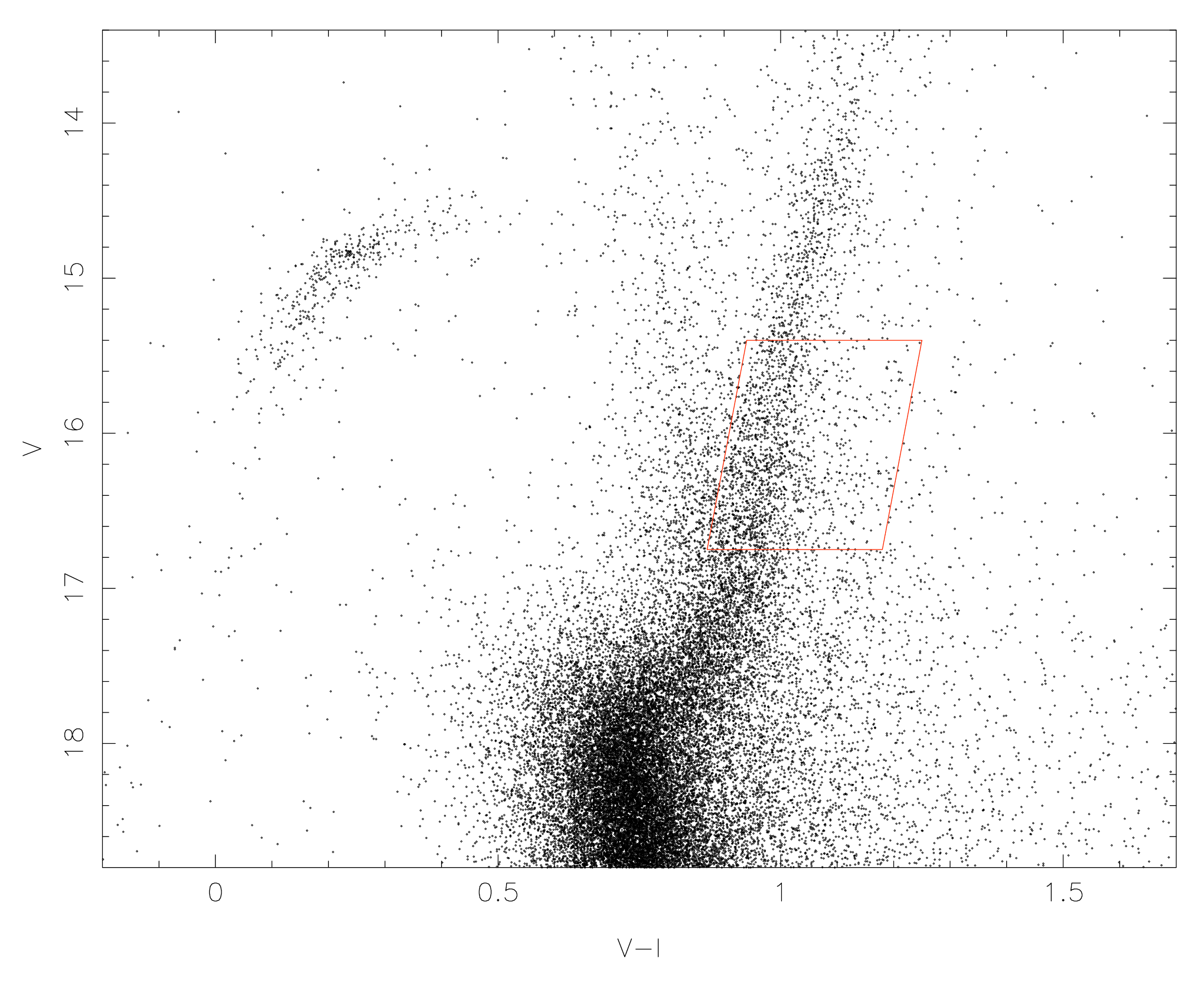}
\caption{A color-magnitude for stars with radial distances from the center of \wcen\/ in the
range 10$\arcmin$ to  25$\arcmin$ taken from the photometry of \citet{MC07}.  
The region outlined on the cluster giant branch defines the selection box for the candidate
sample observed spectroscopically. \label{cmd}}
\end{figure}

In a forthcoming paper, \citet{MC07} will present the results of an imaging survey of a region 
$\sim$2.4 $\times$ 3.9  deg$^2$ in size centered on \wcen.  The $V$ and $I$-band photometry 
was obtained with the 8k $\times$ 8k CCD mosaic camera mounted on the ANU 1m telescope at 
Siding Spring Observatory.  The observation, reduction and analysis techniques employed are 
essentially the same as those in \cite{MC05A} and \cite{MC05B}.
Fig.\ \ref{cmd} shows a color-magnitude diagram (CMD) from the survey for stars that lie
in the radial range 10$\arcmin$ $\leq$ r $\leq$ 25$\arcmin$ from the cluster center.  The principal cluster
sequences such as the horizontal and giant branches, and the main sequence turnoff, are readily
visible although equally, considerable field contamination is visible, a consequence of \wcen's
relatively low galactic latitude (b $\approx$ +15$\arcdeg$).  It is fortunate, however, that the radial 
velocity of the cluster is relatively large (+232 \kms) permitting a generally clean separation 
of members from non-members via the measurement of radial velocities \citep[e.g.][]{LS06}.

A substantial radial velocity survey is required, nevertheless, in order to generate a sample of likely 
members in the outer parts of the cluster and beyond.  The two-degree field (2dF) multi-fibre
instrument on the Anglo-Australian Telescope \citep{LW02} provides an efficient way to conduct
such a survey.  The 2dF is a double-buffered system allowing the fibres for the next field to be 
configured while observing the current field.  The reconfiguration time is of order 45-60 minutes.  
Consequently, the most efficient survey rate is achieved by selecting a magnitude interval that allows 
spectra with appropriate signal-to-noise for velocity and line-strength measurements to be obtained 
in total exposure times comparable to the reconfiguration time, for a given 2dF instrumental setup.
The instrumental setup chosen used the 1200B gratings centered at $\lambda$4250\AA\/ giving
a wavelength coverage of approximately $\lambda\lambda$3700-4800\AA\/ at a scale of 
1.1 \AA/pix.  The resulting spectra have a resolution of $\sim$2.5\AA.  These considerations
led to the choice of the magnitude range 15.4 $\leq$ $V$ $\leq$ 16.75 for the initial sample selection.
The corresponding color limits were then set by allowing for photometric errors and by noting that the
giant branch of \wcen\/ has a significant intrinsic color width.  The selection box is outlined on
Fig.\  \ref{cmd}.  We note that over the area of the imaging survey there are significant variations
in the amount of interstellar reddening \citep[cf.][]{LMC00, MC07}.  Consequently, the limits of the 
selection box were altered by the appropriate amounts given the difference in mean reddening between
each CCD mosaic field and the central field.  The largest shift applied was 0.08 mag in $V-I$ and 0.18 
in $V$.  The initial selected sample was then modified to exclude any candidate that lay within 
$\sim$10$\arcsec$ of a bright neighbor.

The imaging survey covers a rectangular region $\sim$2.4$\arcdeg$ (EW) by 3.9$\arcdeg$ (NS)
in size while the 2dF field is circular.  We therefore chose four 2dF field centers to the N, S, E and W of
the cluster center to maximize the coverage of the imaging survey area.  Stars closer to the cluster
center than 20$\arcmin$ were excluded to minimize crowding concerns while still leaving enough
likely members to define radial velocity and line strength distributions.  Six configurations were generated for both the N and S field centers and four for the E and W fields with a total of 
approximately 5500 candidates.  For each field approximately 15 stars were kept in common between 
all configurations while a modest number of stars in the overlap regions between fields 
were retained in the configurations for both field centers, all to allow monitoring of any systematic effects.
Each configuration also included $\sim$40 fibres allocated to blank sky regions, evenly distributed
between the two 2dF spectrographs.  These regions were selected from the imaging survey to
contain no detected stars (i.e.\ $V \gtrsim 20$) within a radius of 10$\arcsec$.

\subsection{Observations}

The observing run took place over 5 nights in 2003 May and a total of 16 configurations were
observed: six in each of the N and S fields and two each in the E and W fields.  The observing
sequence for each configuration consisted of three 1800 sec exposures preceded and followed
by arc lamp exposures and a fibre flat field exposure.   Multiple exposures,
including exposures recorded in each of the two 2dF spectrographs,
were also obtained of six bright radial velocity standards.  These spectra provide
templates for cross-correlations with the program star spectra and for setting the velocity zero point.

\begin{figure}
\epsscale{0.76}
\plotone{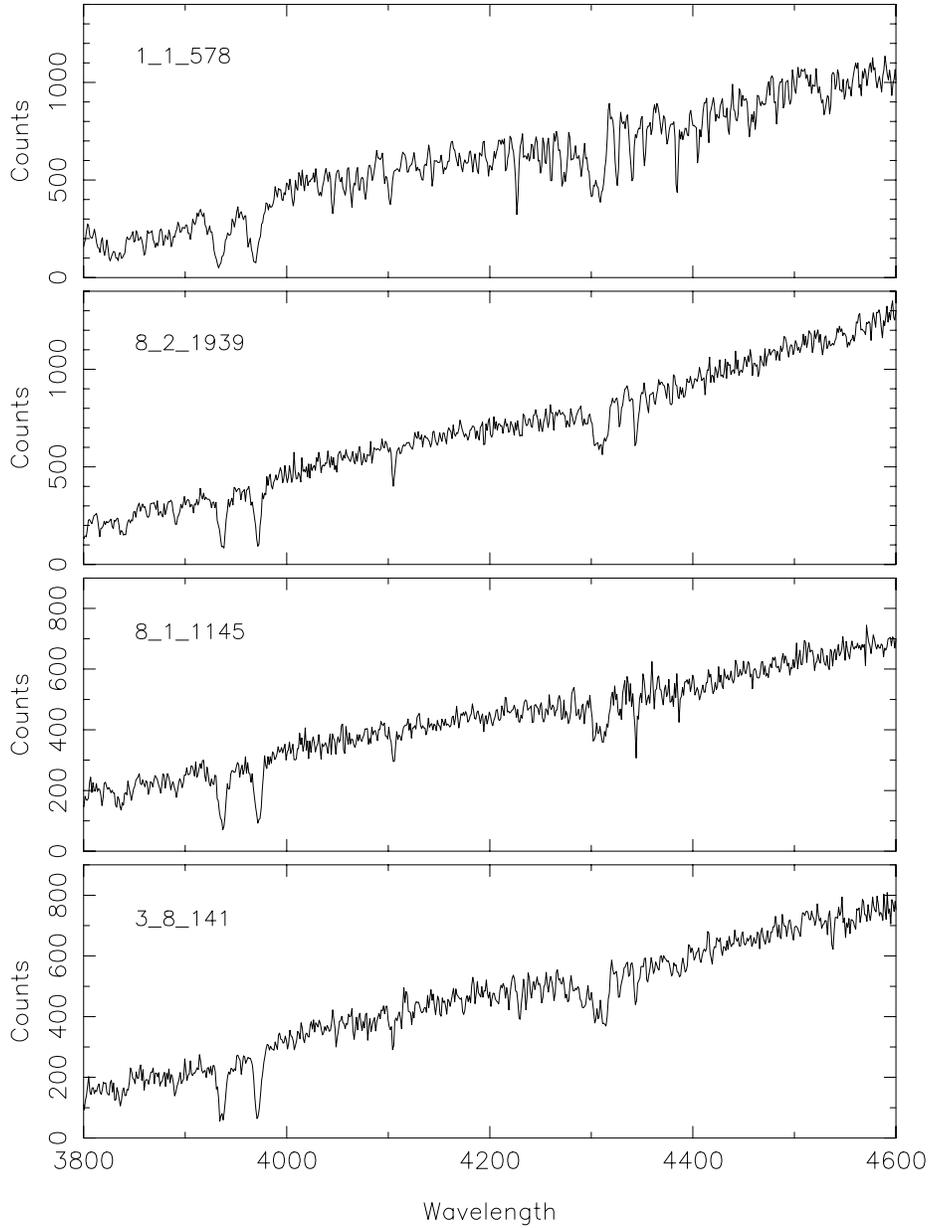}
\caption{Examples of the reduced spectra.  The uppermost spectrum is typical of the low velocity 
relatively metal-rich field star contaminants.  The middle two spectra have radial velocities 
compatible with \wcen\/ membership and the stars lie well inside the cluster tidal radius.  The
lowermost spectrum is for a star with a radial velocity compatible with that for the cluster but which
lies well outside the tidal radius. \label{spectra}}
\end{figure}

The observed data sets, which come in pairs as the 400 2dF fibres feed two separate spectrographs
each imaging 200 fibres onto a CCD detector, were reduced with the 2dF reduction program {\it 2dfdr}.  The fibre flat field exposure was used to generate the `tram line' map, which was then rotated
and shifted to track the pixel locations of the individual fibre spectra.  The fibre spectra were then
`fit' extracted with the wavelength calibration coming from the appropriate arc lamp exposures.  The
fibre spectra for the blank sky regions were then median-combined to form a single sky spectrum,
which was then subtracted from each object spectrum using a weighting appropriate for the 
transmission of the object fibre.  The relative fibre transmissions were determined either from
twilight sky exposures with the same configuration, or from median combined offset sky exposures
obtained as part of the observing sequence.  The reduced spectra from the individual exposures for 
each configuration were then median combined to remove cosmic-ray contamination.
Examples of the reduced spectra are shown in Fig.\ \ref {spectra}.

\subsection{Radial Velocities} \label{rv_sect}

The radial velocities were determined by cross-correlation techniques using the IRAF routine 
{\it fxcor}.  The velocity analysis was carried out independently for each of the two 2dF
spectrograph data sets, including the use of template spectra observed with the appropriate spectrograph.  Only spectra that had at least 400 counts at the G-band ($\sim$$\lambda$4300\AA) were 
included in order to minimize the velocity errors.  The primary template spectra
employed were those of the $V$=6.20 F8 IV-V star HD162396 which provided the best spectral match
to the program stars.   The correlation was carried out over the wavelength range 
$\lambda\lambda$3800--4700\AA, a spectral region that encompasses a number of strong features such as the Ca{\small II} H and K lines, the G-band of CH and the H$\delta$ and H$\gamma$ 
hydrogen lines (cf.\ Fig.\ \ref{spectra}).  

\begin{figure}
\epsscale{0.65}
\plotone{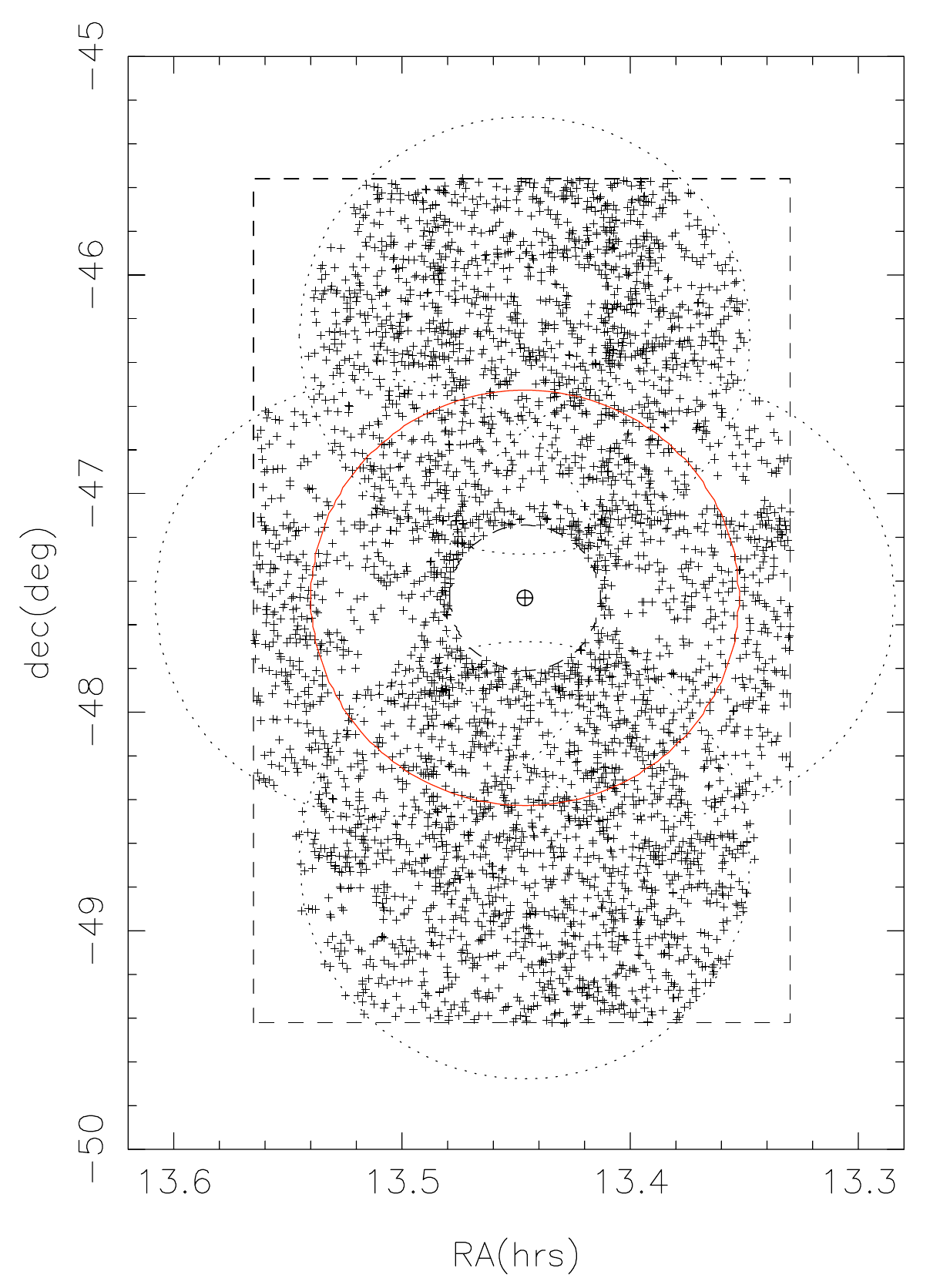}
\caption{The location on the sky of the 4105 program stars with measured radial velocities.  The 
rectangular region outlined with dashed lines is the extent of the imaging survey described
in \citet{MC07} while the four circles marked with dotted lines outline the four 2dF fields.
The center of the cluster is labeled with a circled plus-sign and the red circle marks the adopted
cluster tidal radius of 57$\arcmin$.  No stars were observed inside 20$\arcmin$ from the cluster
center.   Since a velocity was not determined for all possible candidates, some artifacts (mostly
from the fibre allocation process) are visible in the overall distribution. \label{vel_pos}}
\end{figure}

Heliocentric corrections were then applied to correct for the Earth's motion resulting in velocities for 
4105 program objects relative to the velocities for the two template spectra (i.e.\ the velocity zero points
for the two data sets) which remain to be determined.   These were determined through the 
observations of standard stars of known velocity: the zero points are set so that the mean difference
between the observed and catalogue velocities for the standards is zero.  Six observations of four standards in addition to that of the primary template were employed for the spectrograph 1 data set 
while five observations of the same four standards plus that of the primary template were used for the 
spectrograph 2 data.  The uncertainty in the zero points is then the standard error of these mean differences.   The values are $\pm$2 \kms\/ for the spectrograph 1 data and $\pm$3 \kms\/ for 
the spectrograph 2 data.  The overall process was verified by making use of the eight programs stars 
that had at least two velocity determinations in both the spectrograph 1 and the spectrograph 2
data.  The mean velocity difference (spectrograph 1 -- spectrograph 2) for these stars is 2 \kms\/
with a standard error of the mean of 4 \kms, indicating excellent consistency between the
two data sets.  The errors in the individual program star velocities were then determined from the 19 
program stars that had at least five observations during the observing run, including observations across both data sets.  The mean standard deviation from these sets of multiple observations, which we take as the error associated with a single program star velocity determination, is 11 \kms; the uncertainty
in this value is $\pm$1 \kms.  Fig.\ \ref{vel_pos} shows the spatial location of the 4105 program
stars with measured velocities.  

\section{Analysis}

\subsection{Stars within the tidal radius}

\begin{figure}
\epsscale{1.0}
\plotone{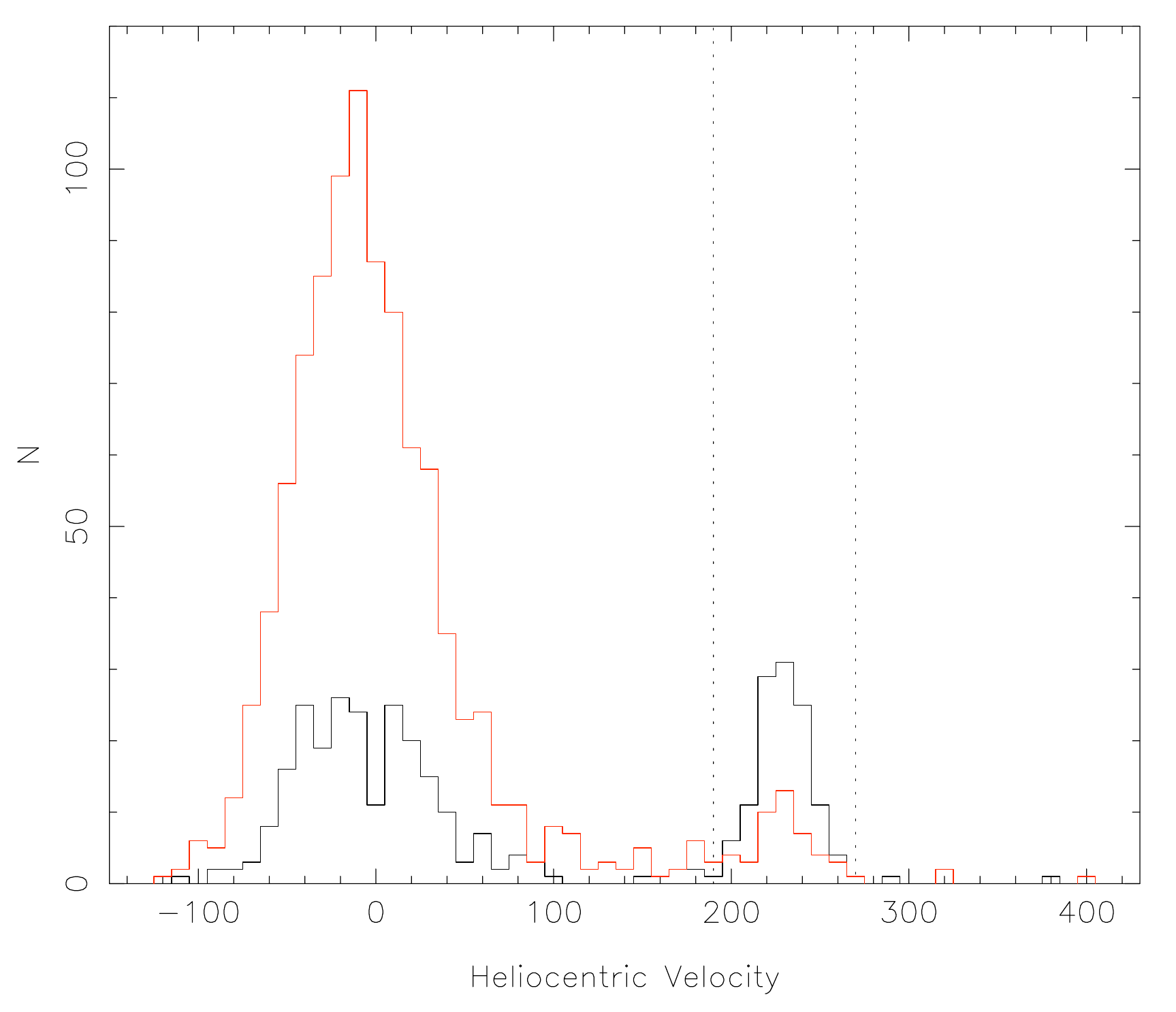}
\caption{Velocity histograms for stars within 57$\arcmin$ of the center of \wcen.  The black line
outlines the velocity distribution for stars with 20$\arcmin \leq r\arcmin \leq 30\arcmin$ and the
red line outlines the distribution for stars with 30$\arcmin \leq r\arcmin \leq 57\arcmin$. 
The systemic velocity of the cluster is 232 \kms\/ and the dotted vertical lines indicate the
velocity interval from which candidate cluster members are selected. \label{vel_histo1}}
\end{figure}

Figure \ref{vel_histo1} shows velocity histograms for all stars inside the adopted cluster
tidal radius of 57$\arcmin$.  This value has been taken from the compilation of \citet{WH96}.  Given
the importance of this parameter, a brief discussion is appropriate.   
\citet{TDK95} list a core radius of 2.6$\arcmin$ for the cluster and a concentration index {\it c}=1.24, which implies a tidal radius of 45$\arcmin$.  \citet{GDC79} derived similar values (2.6$\arcmin$, 1.22, 
44$\arcmin$) from \citet{IK66} model fits, but noted that the observed data indicated a larger 
value for the tidal radius.
The best-fit dynamical models of \citet{GM87} and \cite{MM95}, which are a sizeable improvement over 
the simple \citet{IK66} models, have tidal radii  between 46$\arcmin$ and 60$\arcmin$, with the larger values showing a better fit to the outermost observed surface density data points. 
The data presented in \S 3.3 also imply a tidal radius consistent with the \citet{WH96} value.  
Fortunately, none of the results of this paper depend significantly on 
the actual value of the cluster tidal radius, provided it is similar that adopted to within a few arcmin.
  
The histogram in Fig.\ \ref{vel_histo1} outlined 
by the solid line shows the velocity distribution for 
the stars with radial distances from the cluster center between 20$\arcmin$ and 30$\arcmin$.  It
is clearly bimodal with one population, predominantly foreground disk stars, centered near zero
velocity and a second population, predominantly cluster members, centered near the cluster
velocity of 232 \kms\/ \citep[e.g.][]{vdv06}.  The mean velocity for the 119 stars within the
velocity interval 190--270 \kms\/ (chosen to encompass all likely members given the distribution
in Fig.\ \ref{vel_histo1}) is 229 \kms\/ with a dispersion of 14 \kms.  The mean velocity agrees well 
with the known cluster velocity given the uncertainty in the velocity system zero point discussed above.
Moreover, the observed dispersion value is consistent with that expected from the combination of the intrinsic dispersion of the cluster at these radii, $\sim$8 \kms, \citep[cf.][]{vdv06} and the individual 
velocity errors.  These stars will be referred to as the {\it inner} cluster sample.

The histogram outlined by the dot-dash line in Fig.\ \ref{vel_histo1} is for stars with radial distances
between 30$\arcmin$ and 57$\arcmin$.   Because of the larger area and the decreasing density of
cluster stars with increasing radial distance, the contaminating population now considerably outweighs
the cluster population, which nevertheless remains detectable above the high velocity tail
of the field star distribution.  Using the same selection interval as for the inner region, the 46
candidate cluster members have a mean velocity of 227 \kms\/ and a velocity dispersion of 17 \kms.
The increased velocity dispersion over that seen for the inner region presumably suggests that 
this outer sample is contaminated to some extent with non-cluster members\footnote{We have been
allocated additional time on the Anglo-Australian Telescope 
to measure more precise velocities for stars in this sample, in order to improve the constraints on 
the velocity dispersion in the outer parts of the cluster.}. These stars will be referred to as the 
{\it outer} cluster sample.

Two other criteria can be used to investigate the question of cluster membership.  First, noting
that the expected proper motion of cluster members is small \citep[cf.][]{D99a}, we can
use published proper motion catalogs to weed out from the sample any stars with significant motions,
as they are unlikely to be cluster members (though equally a small proper motion does not
guarantee cluster membership).  Second, while it is well established that \wcen\/ has a substantial
internal abundance range, we can still use the spectra to define relations between various line
strength parameters and again seek outliers from the relations defined by the majority of stars.

Considering first proper motions, since the majority of the candidates are too faint to be
included in the UCAC2 catalog \citep{Z04}, we extracted from the SuperCOSMOS Science 
Archive\footnote{ http://surveys.roe.ac.uk/ssa/index.html} proper motions for all the stars in the 
outer cluster sample 
(due to crowding effects we did not consider the proper motions for the inner cluster sample).  
\cite{H01} indicate that for the magnitude range of our program stars, the estimated proper motion
accuracy is of order 10 mas/yr.  The 
mean motions in RA and Dec for the 46 stars were calculated along with the standard deviations,
which were 6.3 mas/yr in RA and 5.5 mas/yr in dec, respectively.
One star, 6\_6\_1259 (cf.\ Table \ref{nonmembers}) lay outside the error ellipse defined by 
3$\sigma_{\mu,\alpha}$ and 3$\sigma_{\mu,\delta}$.
It was not considered in the analysis of spectral line strengths.

For line strength parameters we first removed the overall continuum shape of the spectra by fitting
a lower order polynomial using the IRAF routine {\it continuum}.  Following, for example, \citet{B99},
we then measured via numerical 
integration the equivalent widths of the Ca{\small II} K line and the G-band of CH and, via gaussian fits, 
the equivalent widths of the H$\delta$ and H$\gamma$ hydrogen lines (cf.\ Fig.\  \ref{spectra}).  
The continuum and feature band passes used are listed in Table \ref{table1}.   The K-line and 
G-band strengths serve as metal abundance indicators while the hydrogen line strengths serve
as an indicator of effective temperature.  The mean uncertainties
in the line strengths were then determined from the standard deviations of the measurements for
the seven stars in the inner sample that have at least 6 observations.  For stars with multiple spectra
the individual line strength measurements were averaged.  Only for one star was it not possible to 
measure all 4 indices: for star 8\_6\_15256 the H$\delta$ line was significantly affected by noise 
and no reliable measurement could be made.  The other line strengths, however, are consistent with
cluster membership.

\begin{figure}
\epsscale{1.0}
\plotone{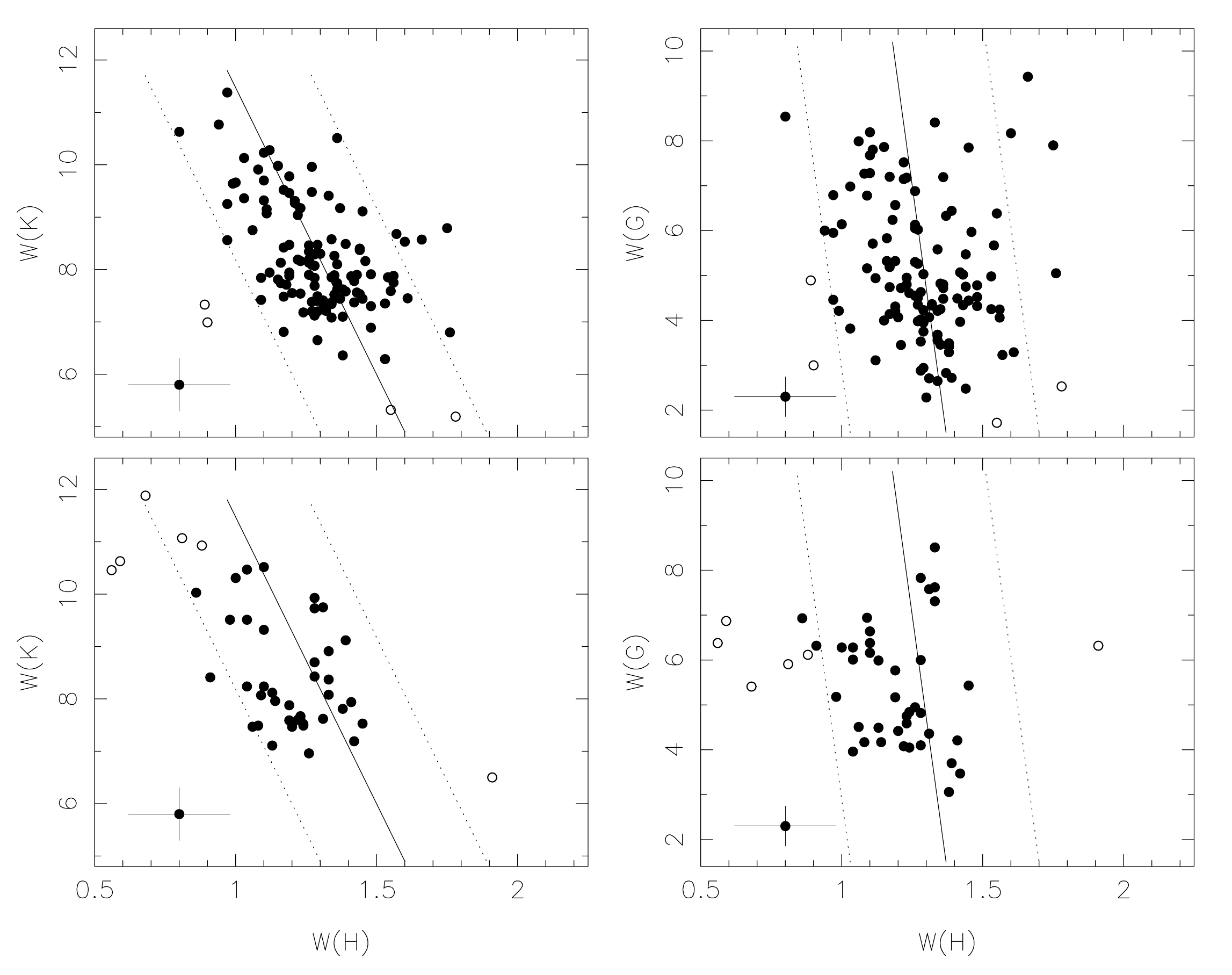}
\caption{Line strength measurements.  The left panels show the equivalent width of the Ca{\small II}
K line plotted against the mean equivalent width of the H$\delta$ and H$\gamma$ hydrogen lines.
The right panels plot the G-band (CH) strength against hydrogen line strength.  The upper panels are 
for the inner sample while the lower panels are for the outer sample.  Open circles represent stars that 
are classified as probable non-members of the cluster.  The solid lines are least squares fits to the
inner sample, with the dotted lines representing  $\pm$2$\times$ the rms about the fit .  The fits
are reproduced in the lower panels.  Mean error bars are shown in the bottom left corner of each panel.  \label{ew_fig}}
\end{figure}

\begin{figure}
\epsscale{0.72}
\plotone{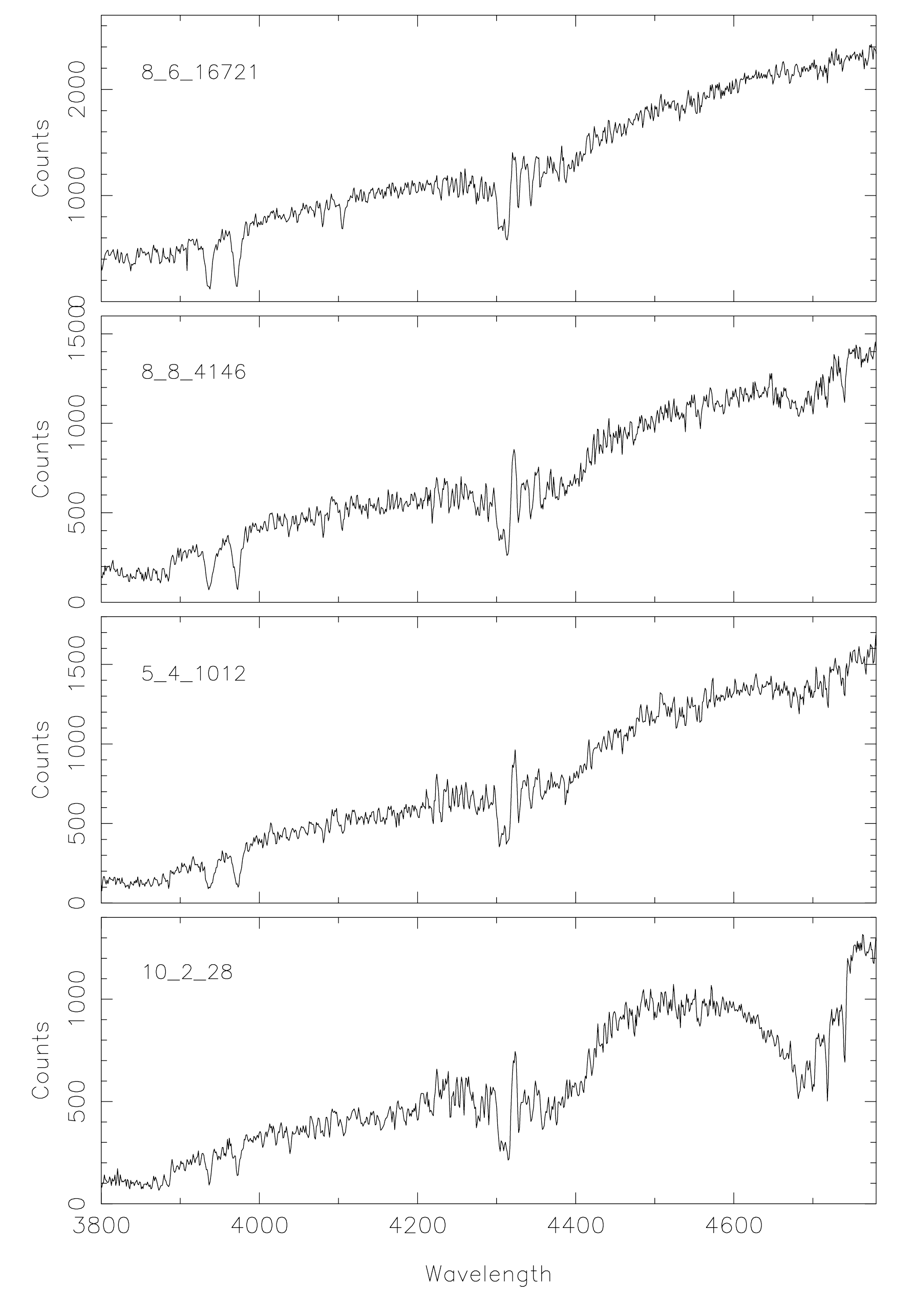}
\caption{Spectra of stars showing enhanced carbon features at the G-band, $\lambda \approx 4300$\AA,
at additional A-X CH bands, $\lambda \approx 4370$\AA, and at C$_{2}$ bands, $\lambda \approx 
4700$\AA.  The upper two stars come from the inner sample while the third is from the outer
sample.  The star in the bottom panel is clearly a carbon star.  Its membership status is discussed in 
the text.  \label{Cstarsfig}}
\end{figure}

Given that the inner sample is likely to consist largely of cluster members, we used this sample
to define relations between the K-line strength and the average hydrogen line strength 
$W_{H}=[W(H)_{\delta}+W(H_{\gamma})]/2$, and between the G-band strength and $W_{H}$ for 
probable cluster stars.  These relations are shown in the upper panels of Fig.\ \ref{ew_fig}.  
Turning first to the upper left panel, we can see that, as expected given the well established 
abundance range in \wcen, there is a significant range in 
the K-line strengths for the inner sample stars that is anti-correlated with hydrogen line strength.  
The solid line in the panel represents a least 
squares fit to the data, while the dotted lines represent $\pm$2$\times$ the rms about the fit after excluding the four stars plotted as open circles.  Two of these are clearly deviant, but exclusion of
the two stars with the weakest K-lines deserves some comment.  The abundance distribution in 
\wcen\/ is well established \citep[e.g.][]{NFM96} and is characterized by a steep initial rise to the
mean abundance followed by a long tail to higher abundances.  Excluding the two stars with
the lowest K-line strengths, the distribution of K-line strengths in the panel reflects this 
general shape and for that reason the two stars have been excluded.  A small number of stars also
lie to the right of the +2$\sigma$ line in the panel.  These are nevertheless considered members
because they all show very strong G-bands and other features indicating probable carbon
over-abundances (see Fig.\ \ref{Cstarsfig}).  The exclusion of 4 stars from the inner sample
as possible non-members is 
consistent with the number of field stars expected in the adopted cluster velocity range, given the
number of stars in the velocity intervals adjacent to that for the cluster seen in the solid histogram
in Fig.\ \ref{vel_histo1}.

The upper right panel of Fig.\ \ref{ew_fig} shows the distribution of G-band strengths for the inner 
sample.  Once again the solid line is a least squares fit and the dotted lines are $\pm$2$\times$ the 
rms about the fit after again excluding the four stars plotted as open circles.  There is clearly a 
very large range in G-band strengths for the stars in this sample.  Many of the stars with strong
G-bands also show additional CH absorption in the region around $\lambda$4370\AA; some examples are shown in Fig.\ \ref{Cstarsfig}.  The existence of stars in \wcen\/ with a range of overabundances
of carbon is well established; see, for example, \citet{LDN07} and the references therein.

The lower left panel shows the W$_K$,W$_H$ for the stars in the outer sample.  The solid and
dotted lines are reproduced from the equivalent panel for the inner sample.  Given that there is 
evidence for a radial abundance gradient in the cluster, in the sense that metal-rich objects are
less frequent in the outer regions \citep[e.g.][]{NFM96}, it seems unlikely that the outer sample would
contain comparatively more metal-rich objects than the inner sample.  Consequently, for the
stars with strong K-lines, we have considered stars with W$_K$ $\gtrsim$ 10.6 \AA\/ to be 
non-members even if they are within the $\pm$2$\sigma$ boundaries.  In many cases these
stars have G-band strengths that fall outside in the $\pm$2$\sigma$ boundary in the lower
right panel, confirming their probable non-member status.  All-in-all we classify 6 of the outer sample 
stars as non-members.  As noted above there is seventh probable non-member in the outer
sample based on a discrepant proper motion.  Interestingly, like the inner region, the  outer region 
also contains a number of CH-strong stars.  Examples are shown in Fig.\ \ref{Cstarsfig}.   Inspection
of the dashed line histogram in Fig.\ \ref{vel_histo1} suggests that excluding 7 stars from the outer
sample as likely non-members is perhaps a conservative number; there may well be an additional
number of approximately the same size of non-member stars in this sample that we have not been
able to conclusively identify.  Such identification would require velocities with a higher degree of
precision than that presented here as well as knowledge of the cluster velocity dispersion in this outer
region.

One star deserves special mention.  This is star 10\_2\_28 whose spectrum is shown in the bottom
panel of Fig.\ \ref{Cstarsfig}.  With strong C$_{2}$ features at $\lambda\approx$ 4700\AA, strong 
CH absorption features in the region $\lambda \lambda$4250--4430\AA\/ and strong CN
absorption features at  $\lambda\approx$ 3883\AA\/ and  $\lambda\approx$ 4215\AA, this star is 
clearly a carbon-star.  Such stars are not unknown in \wcen\/ \citep[e.g.][]{vL07} but the known
members stars with similar spectra, such as ROA 577 \citep[see Fig.\ 25 of ][]{vL07}, are found on the 
red giant branch at considerably brighter magnitudes.  Determining the radial velocity of 10\_2\_28
is not straightforward given the very different spectral type.  Nevertheless, cross-correlations with
the latest spectral-type among the observed radial velocity standards and with probable \wcen\/
members such as 8\_8\_4146 (cf.\ Fig.\ \ref{Cstarsfig}) indicate a velocity of $\sim$260 $\pm$ 15
\kms.  This is nominally within the velocity range of probable cluster members but the star lies
at a distance of 47.3$\arcmin$ from the cluster center.  Given that the number of cluster members
at this radius is small, and that cluster members of this type are comparatively rare, we conclude
that this star is likely not a member of the cluster.  Interestingly there is a second carbon star
that lies in the field of the cluster but which is clearly not a cluster member
\citep[ROA 153, see][]{SW73}.

In Table \ref{members} we list the ID number, the radial distance from the cluster center, the 
right ascension and declination, the number of observations, the radial velocity and the $V$ and 
$V-I$ photometry \citep[from][]{MC07} for the 154 probable \wcen\/ members categorized in this work. 
At a distance of 54.7$\arcmin$ from the cluster center, star 9\_4\_1918 is the most radially
distant probable cluster member inside the adopted tidal radius; there are 9 such stars beyond 
40$\arcmin$ from the cluster center.  For completeness in Table \ref{nonmembers} we list the 
same information for the 12 stars in the inner and outer cluster samples that are classified as 
probable non-members, including the carbon star 10\_2\_28, despite having radial velocities
comparable to that of the cluster.   Similarly, Table \ref{vel_nonmems} 
lists the same information for the 1183 stars inside the adopted cluster tidal radius that have velocities
outside the velocity interval (190--270 \kms) used to select cluster member candidates.
 
\begin{figure}
\epsscale{0.7}
\plotone{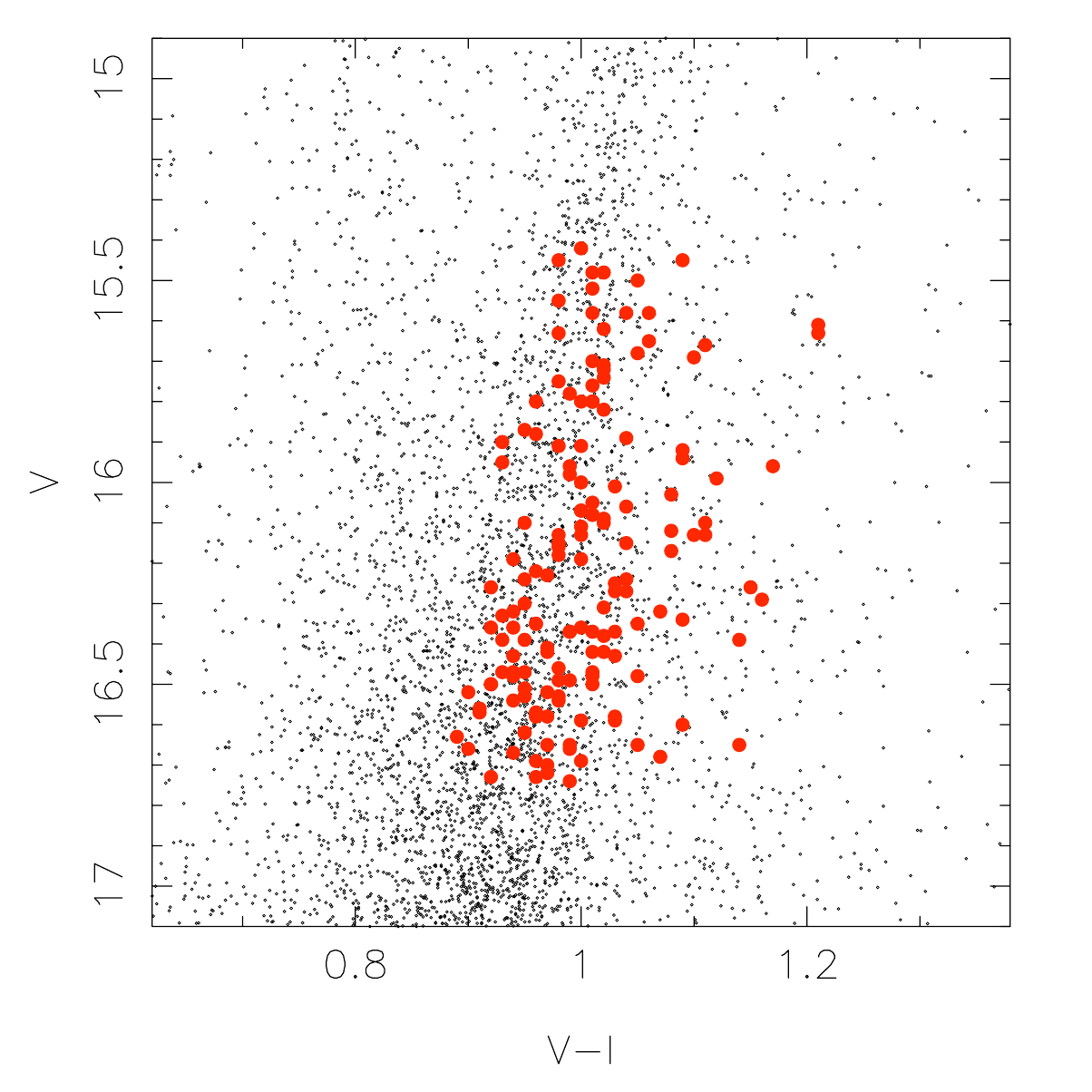}
\caption{The $V,I$ photometry for the 154 probable members of \wcen\/
superposed on the photometry from Fig.\ \ref{cmd}.      \label{2cmd_fig}}
\end{figure}

Fig.\ \ref{2cmd_fig}
shows the photometry for the probable members superposed on the photometry for all
stars in the 10$\arcmin$--25$\arcmin$ radius region (cf.\ Fig.\ \ref{cmd}).  As expected given
the cluster metallicity distribution, the blue side of the color distribution of the probable members
is sharply defined but there is a considerable spread to redder colors.  

\subsection{Stars beyond the tidal radius}

\begin{figure}
\epsscale{0.95}
\plotone{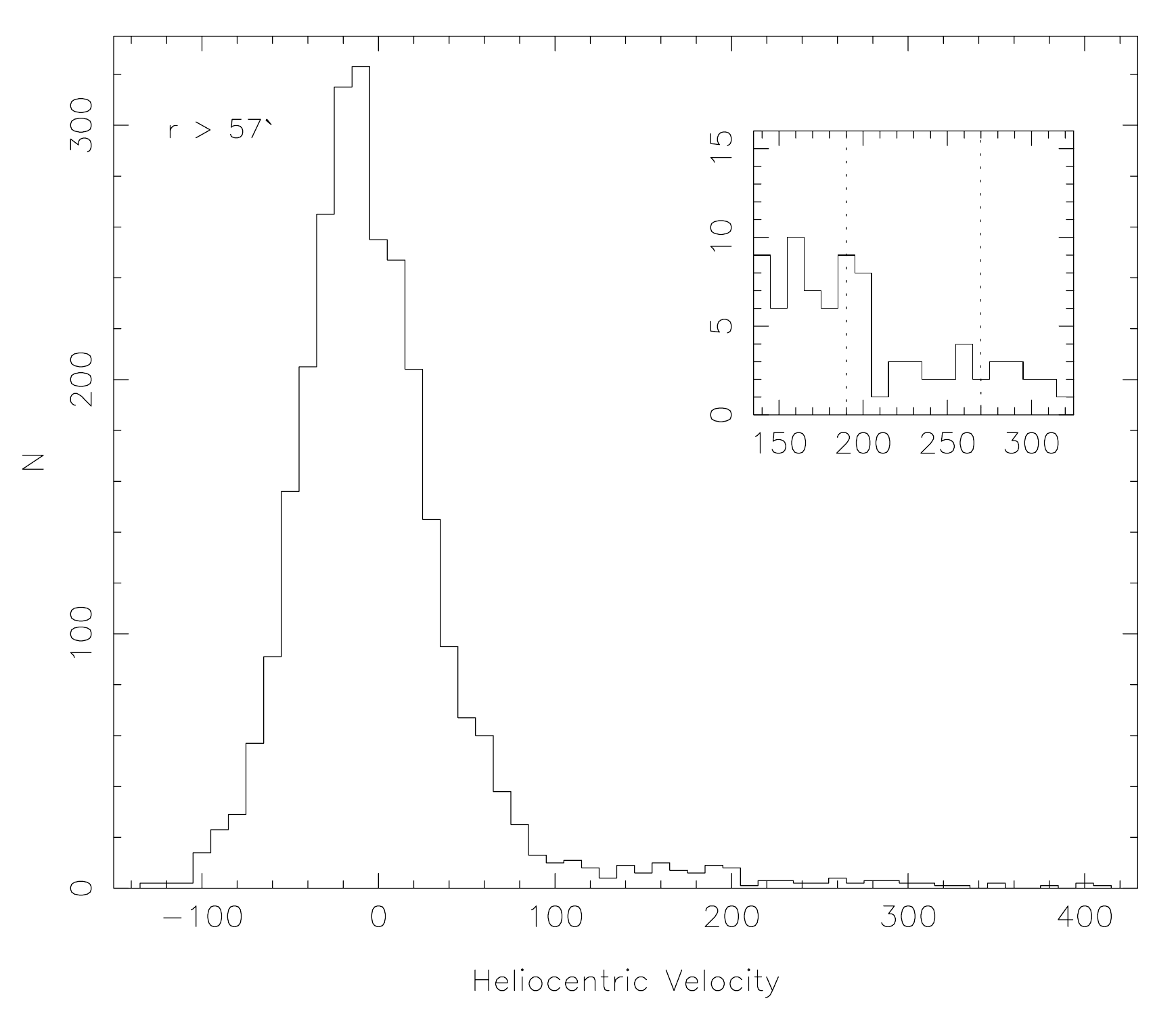}
\caption{Velocity histogram for stars lying further than 57$\arcmin$, the adopted tidal radius,
from the center of \wcen.  The insert panel shows the distribution near the heliocentric velocity 
of the cluster and the dotted lines outline a $\pm$40\kms\/ window about that value (cf.\ Fig.\ \ref{vel_histo1}).   \label{vel_histo2}}
\end{figure}

In Fig.\ \ref{vel_histo2} we show the velocity distribution for stars beyond the adopted tidal radius.
The distribution is clearly dominated by disk stars which have heliocentric velocities near zero (for
reference, zero velocity in the galactic rest frame at the Sun's location corresponds to a heliocentric 
velocity of 168 \kms\/ in the direction of \wcen, assuming a galactic rotation velocity of 220 \kms).  
The insert in the figure shows a close-up of the distribution in the vicinity of the velocity of the cluster.  
While the numbers of stars are small and therefore subject to significant statistical fluctuation, there is 
no obvious indication of an excess of stars at or near the cluster velocity.  

Nevertheless there are 28 stars in the $\pm$40 \kms\/ velocity range shown in the insert frame in 
Fig. \ref{vel_histo2}.  Of these, two have proper motions inconsistent with association with \wcen\/ and
all (including the stars with discrepant proper motions) but 9 occupy locations in the 
(W$_K$,W$_H$) and (W$_G$,W$_H$) diagrams that are inconsistent with \wcen\/ membership 
(cf.\ Fig.\ \ref{ew_fig}).  Most of these likely non-member stars have relatively
large values of W$_K$; 18 have W$_K$ $\geq$ 9.5\AA\/ (cf.\ Fig.\ \ref{ew_fig}).
They are most likely members of the Galaxy's bulge population lying at distances beyond that
of \wcen.  The line-of-sight through the cluster has a minimum distance of 10.4 kpc from the Galactic
Center at a height of 3.4 kpc above the Galactic plane.  At this distance the stars would have 
M$_V$ $\approx$ +0.5, approximately the magnitude of the red clump, which is not inconsistent with 
the proposed interpretation.

For the 9 stars with line strengths consistent with cluster membership, two have 
velocities of 192 and 195 \kms, respectively, while a third has V$_{r}$ = 261 \kms.  The remaining six 
have a mean velocity of 233 \kms, corresponding closely to that for
the cluster, and a dispersion of only 9 \kms, which given the velocity errors, is consistent with 
an intrinsic dispersion of near zero.  These are exactly the characteristics expected for an extra-tidal 
population and we tentatively identify these six stars {\it as likely extra-tidal (former) cluster members}.
Table \ref{tab_4} lists their properties.  The spectrum of one of the stars is shown in the bottom
panel of Fig.\ \ref{spectra}.  Given the small number and the shape of the region sampled
(cf.\ Fig.\ \ref{vel_pos}), no inferences can be made from the spatial distribution of these stars, except
to note that the furthest lies almost exactly twice the adopted tidal radius from the cluster center (which 
is also the approximate limit of the area surveyed) and
that the surface density corresponds to 1.2 stars per square degree for the region between one and 
two tidal radii.  Clearly more accurate velocities and abundance information are required to more firmly
establish the association or otherwise of these stars with \wcen, but for the present we will assume that
they are all extra-tidal members of the cluster and treat derived quantities as upper limits.  The
alternative assumption, that all or some of these stars are simply members of the field halo
population, cannot be ruled out.

The immediate question to ask is what fraction of the mass of \wcen\/ is potentially represented by the
extra-tidal stars.  Assuming that the extra-tidal population is identical with that of the cluster,
we can estimate the fraction as follows.   Using a core radius of 2.4$\arcmin$ and a tidal radius of 
57$\arcmin$ and
the integral formulae given in \cite{IK62}, the percentage of the total luminosity of the cluster contained
between 20$\arcmin$ and the tidal radius is approximately 10 per cent.  Assuming mass follows
light, this figure also represents the fraction of the total mass contained in this radial region.  The
region is also characterized by the 154 likely members we have identified, whereas using the 
surface density calculated above, the region between one and two tidal radii is estimated to
contain, {\it at most}, approximately 10 extra-tidal members with the same characteristics.
Consequently, combining these numbers, we conclude that an {\it upper limit} on the amount of mass 
in extra-tidal material lying
between one and two tidal radii from the cluster centre is of order 0.7 $\pm$ 0.2 per cent of the total cluster mass.   It may well be considerably less.  The uncertainty quoted
comes solely from assuming Poisson statistics apply to the sample
of 6 identified extra-tidal member candidates.   This limit is comparable to the results of \citet{LK07}
who estimated a $\sim$1\% upper limit on the extra-tidal populations of three globular clusters
(NGC~288, M30 and M55) but differs from that of \citet{GJ06} who estimate that at least 3\% of 
the present cluster mass is associated with the tidal tails of the globular cluster NGC~5466.  
The biggest contrast is of course with Pal 5, where the mass in the extensive tidal tails exceeds the 
mass of the cluster itself \citep[e.g.][]{MO03}.

\subsection{The surface density profile}
 
\begin{figure}
\epsscale{0.55}
\plotone{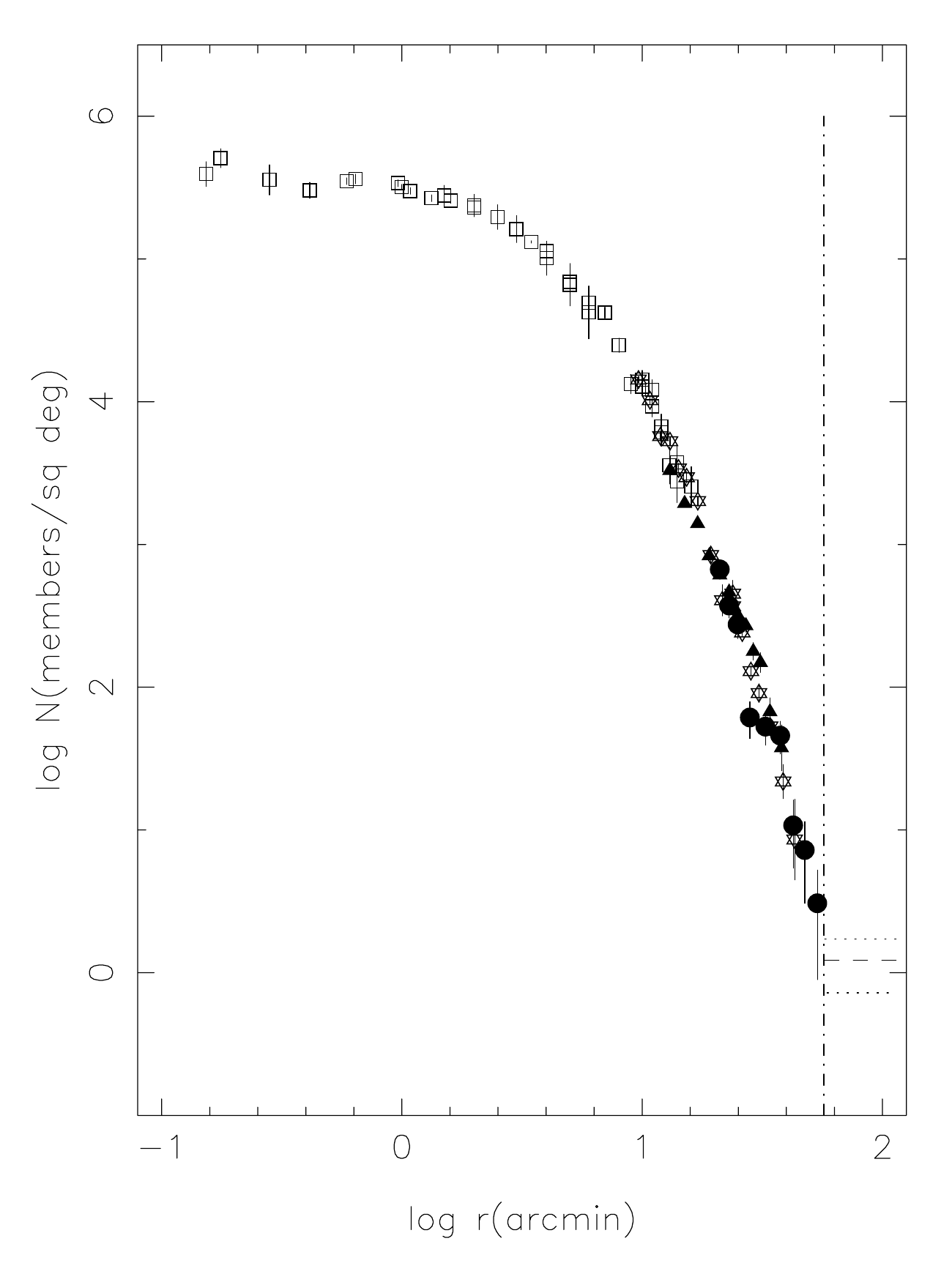}
\caption{The (circularly averaged) surface density profile for \wcen.  Open squares and star symbols 
are surface photometry 
and star-count data, respectively, from \citet{GDC79}.  Filled triangles are data from \citet{MC07}
while the filled circles are derived from the cluster member sample determined in this paper.  The
vertical scale corresponds to the surface density of stars in the cluster member sample.  It can be 
converted to the surface brightness scale of \citet{GDC79} by subtracting 8.318.
The vertical dot-dash line indicates the adopted tidal radius.  The horizontal dashed line shows the
surface density inferred for the region between 1 and 2 tidal radii if the six candidate extra-tidal stars 
are indeed associated with the cluster. 
The dotted lines represent the statistical uncertainty in that density.   \label{sb_prof}}
\end{figure}

In Fig.\ \ref{sb_prof} we show the (circularly averaged) surface density profile of \wcen\/ from the 
center to the outermost regions, covering a factor of $\sim$10$^5$ in surface density.  The profile combines the surface photometry and star-count data from \citet{GDC79}\footnote{These data differ 
insignificantly from those of \citet{GM87} who adopted a very slightly different normalization of the 
star count data to the surface photometry.}, together with the circularly averaged surface density data 
from \citet{MC07} and that derived from our cluster member sample.   The points from \citet{MC07} are
for stars from just below the turnoff and brighter (cf.\ Fig.\  \ref{cmd}) and are for regions where
image crowding is not a concern.  The background density was determined from the extensive region 
beyond the tidal radius covered in the imaging survey.   The stars included correspond to those that 
dominate the integrated light and thus no systematic effects are expected in scaling the surface
density count data to the surface brightness observations \citep[cf.][]{GDC79} .  In any case the long 
2-body relaxation time of \wcen\/ \citep[e.g.][]{WH96}, especially in the outer parts, means that significant 
mass segregation in this cluster is unlikely.  Indeed \citet{vdv06} find no dynamical evidence for any 
change in {\it M/L} with radius, as expected given the relatively long relaxation time. 

The surface density points from the cluster member sample were determined as follows.
First, we grouped the radial distribution of the 154 probable \wcen\/ members into 
annuli.  Mean surface densities were then calculated with errors determined from the number of 
stars in each annulus.  While Fig.\ \ref{vel_pos} shows that each annulus has not been 
completely surveyed,
we assume that there is no radial variation in the degree of incompleteness.  The resulting surface densities were then matched with the existing data in the region between 20$\arcmin$ and 26$\arcmin$.  
As is evident from Fig.\ \ref{sb_prof} the agreement between the new and existing data at these, and 
larger radii, is excellent.  This serves as confirmation that our sample, while not complete, is
suitably representative.  Overall the profile shows a rapidly 
declining surface density distribution with a limiting radius consistent with the value 
we have adopted for the cluster tidal radius, shown by the vertical line in the Figure.  
Fig.\ \ref{sb_prof} also shows the density inferred for the extra-tidal population in the region between 
one and two tidal radii, {\it if} the six candidates identified here are all associated with the cluster.  

Given the extensive range in surface density, the complete profile shown in Fig.\ \ref{sb_prof},
not surprisingly, cannot be adequately represented by analytic formulae such as those of 
\citet{IK62}, or by simple models \citep[e.g.][]{IK66}.  However, more complex models, for example those 
of \citet{GM87}, \citet{MM95} and \citet{MM97}, do appear capable of fitting the entire profile.  
We note in passing that the models of  \citet{vdv06} adopt a 
tidal radius of 45$\arcmin$, which on the basis of Fig. \ref{sb_prof}, is somewhat too small.  
The core radius of the cluster is well established at r$_{c} \approx 2.5\arcmin$ \citep[e.g.][]{IK62, GDC79, TDK95}.  Table \ref{sb_tab} lists the surface density data from \citet{MC07}, and from this work, in a 
manner that is directly compatible with Table 1 of \citet{GM87}.
 
\section{Discussion}

The photometric work of \citet{Law03} \citep[see also][]{MC07} concluded that \wcen\/ lacks any 
notable extra-tidal structure, with the extensive tidal tails originally proposed by \citet{LMC00} arising 
from the lack of allowance for the effects of reddening variations across the field (a possibility that 
\citet{LMC00} themselves acknowledged).  Our spectroscopy based results are consistent with, and
strengthen, this result.

As noted in \S1, the presence of extra-tidal structure, particularly in the form of tidal tails, is relatively 
well established in a number of globular clusters \citep[e.g.][and the references therein]
{G95,LMC00,GJ06} with
the most striking example being Pal~5 \citep{MO03, GD06}.  Clearly, however, \wcen\/ does not
obviously fall into this category.   The most likely explanation for this is as follows.  The largest 
dynamical influence on \wcen\/ is the `tidal shocking' that occurs each time the cluster passes
through the Galactic plane \citep[e.g.][]{vdv06}, as is also the case for Pal~5 \citep[cf.][]{DH04}. 
\citet{vdv06} present a detailed model of the structure of \wcen, constrained by high quality
proper motions and radial velocities for many thousands of cluster member stars, and we can
use this model to estimate the effects of tidal shocks on the outer parts of the cluster.
With the parameters given in \S9.5 of \cite{vdv06}, which rely on the \wcen\/ orbital parameters listed 
by \citet{D99b}, the average change in the velocity of an \wcen\/ star as a result of the impulse that
occurs as the cluster traverses the Galactic plane is estimated as
$\mid$$\Delta${\it v}$\mid$ $ \approx$ 0.17 r$\arcmin$ \kms, 
where r$\arcmin$ is the radial distance of the star from the cluster center in arcmin.  
This is to be compared with the
local escape velocity, which can be approximated by twice the local velocity dispersion
\citep[e.g.][p.\ 490]{BT87}.  The form
of the velocity dispersion in the outer parts of \wcen\/ is not well established (even the observations
of \citet{Sc03} reach only a maximum of $\sim$30$\arcmin$ from the center) but it is clear that it is only
in the outermost parts of the cluster that the tidal shock effects are large enough to add sufficient
energy to (some) cluster stars to cause them to escape.  Indeed as \citet{vdv06} note, the phase-space
structure of the outer parts of \wcen\/ may well be completely dominated by the effects of the 
relatively frequent tidal shocks: \citet{vdv06} estimate that the cluster is immersed in the disk and 
feels the
additional gravitational field for $\sim$10\% of the $\sim$120 Myr orbital period.  
The main point here though
is that the region of \wcen\/ which is most strongly influenced by the tidal shocks contains only a 
very small fraction of the total mass of the cluster, unlike the situation for Pal~5
\citep{DH04}.  For example, 
only $\sim$1\% of the \wcen\/ mass lies between $\sim$40$\arcmin$ and the
tidal radius.  Consequently, it is likely that tidal shocks are indeed driving the escape of stars 
consistent with the limits on the extra-tidal population observed here.  Further, we note that if stars leave
the cluster with a relative velocity of $\sim$1 \kms, then that
is sufficient for the star to traverse the region between 1 and
2 tidal radii in less than the orbital period of the cluster around the Galaxy.  Thus even if tidal
shocks are driving the escape of a small number of stars, the number remaining close to the
cluster will be small, again consistent with the observed lack of extra-tidal stars in our study.   
Indeed this suggests that if it were possible to isolate putative \wcen\/ main sequence stars at and
beyond the tidal radius against the enormous field contamination, it may well be possible to characterize tidal tails from \wcen\/.  Such discrimination might be possible with GAIA data though many square
degrees of sky would need to be searched.

In the context of the model in which \wcen\/ is the remnant nucleus of a disrupted former dwarf
galaxy, it is clear from the above that any stars which are more loosely bound than those in the 
outer parts of the remnant nucleus/star cluster will be even more strongly affected by the tidal shocks.  
Consequently, they would have been lost from the vicinity of the remnant nucleus/star cluster at 
relatively early stages in the interaction, once the disrupting dwarf galaxy has approached the current 
orbit of  \wcen.  The dynamical models of \citet{BK03} and \citet{TD03} show that this occurs within 
about 3 Gyr of the initiation of the interaction.   Given that any gas in the original dwarf would have 
been used up in star formation \citep[cf.][]{BK03} (and any remaining stripped out) in the first phases 
of the interaction, and that \wcen\/ contains only stars with ages in excess of $\sim$10 Gyr, 
\citep[e.g.][]{SP05, LS06} it is evident that the more loosely bound stars have been lost from
the vicinity of \wcen\/ many Gyr ago, and are now widely dispersed around the Galaxy.   The kinematic 
signatures of local field stars that may have had their origin with the disrupted system that once
contained \wcen\/ have been modeled, and such stars may well have been observationally detected 
\citep[e.g.][]{D02,MCO03,CM04,MN05}.

In this sense the currently disrupting dwarf galaxy Sagittarius and its central massive star cluster 
M54 provide a similar example to the postulated scenario for \wcen, but on a longer timescale -- eventually the Sgr field stars will disperse uniformly around the galaxy but like \wcen, M54 will retain 
its dynamical integrity for a very long time.

\section{Conclusions}

We have carried out an extensive radial velocity and line-strength analysis of observations 
of a large sample of stars that covers the outer parts of the globular cluster \wcen\/ and a
substantial area beyond the cluster's tidal radius.  While the analysis has led to the establishment
of cluster membership for $\sim$150 stars in the outer parts of the cluster, and revealed that the steeply 
declining radial surface density distribution of these stars is consistent with other extant data, the 
survey has
identified only a very small number of objects that can be plausibly classified as extra-tidal cluster
stars.  The upper limit on the amount of mass potentially contained in the region between
1 and 2 tidal radii is less than one percent of the present cluster mass.  We conclude therefore that 
there is no compelling evidence for any significant extra-tidal population associated with \wcen. 

\acknowledgments
The authors are grateful for the excellent support at the telescope provided by the staff of the 
Anglo-Australian Observatory.  The comments of the anonymous referee led to improvements
in the clarity of the paper.  This research has made use of the SIMBAD database, operated at 
CDS, Strasbourg, France.

{\it Facilities:} \facility{AAT (2dF)}, \facility{ANU Siding Spring Observatory 1m telescope}

\begin{deluxetable}{lccc}
\tablewidth{0pt}
\tablecaption{Spectral feature and continuum bandpasses \label{table1}}
\tablecolumns{4}
\tablehead{
\colhead{Feature} & \colhead{Feature Bandpass} & \colhead{Blue Continuum } & 
\colhead{Red Continuum } \\ & \colhead{(\AA)} & \colhead{Bandpass (\AA)} & \colhead{Bandpass (\AA)}}
\startdata
Ca{\small II} K & 3925.--3943. & 3909.--3921. & 4010.--4030.\\
H$\delta$ & 4093.--4109. & 4010.--4030. & 4120.--4131.\\
G-band (CH) & 4294.--4316. & 4230.--4248. & 4415.--4435.\\
H$\gamma$ & 4334.--4347. & 4230.--4248. & 4415.--4435. \\
\enddata
\end{deluxetable}

\begin{deluxetable}{rccccccc}
\tablewidth{0pt}
\tablecaption{Probable cluster member stars \label{members}}
\tablecolumns{8}
\tablehead{
\colhead{ID} & \colhead{R} & \colhead{Velocity} & \colhead{N}  & \colhead{RA} & \colhead{Dec} & 
\colhead{$V$} & \colhead{$V-I$} \\
 & \colhead{(arcmin)} & \colhead{(km s$^{-1}$)} & & \colhead{(J2000)} & \colhead{(J2000)}  &  & }
 \startdata
 8\_4\_3206 & 20.0 & 239 & 1 & 13 27 47.3 & --47 45 44 & 16.42 & 0.97  \\
 8\_7\_15831 & 20.1 & 224 & 1 & 13 25 06.5 & --47 17 42 & 15.80 &  0.96 \\
 8\_6\_16385 & 20.1 & 219 & 1 & 13 24 56.3 & --47 36 35 & 16.39 & 0.93 \\
 8\_8\_3134 & 20.2 & 221 & 2 & 13 25 47.7 & --47 11 00 & 16.63 & 0.89 \\
 8\_6\_15256 & 20.2 & 197 & 1 & 13 25 11.4 & --47 41 03 & 16.66 & 0.90 \\
 8\_8\_2219 & 20.2 & 228 & 1 & 13 26 04.4 & --47 09 40 & 16.42 & 1.02 \\
 8\_4\_4900 & 20.3 & 248 & 1 & 13 27 20.0 & --47 48 03 & 16.23 & 0.97 \\
 8\_2\_1336 & 20.3 & 253 & 1 & 13 28 43.3 & --47 24 38 & 16.19 & 1.00 \\
 7\_3\_261 & 20.3 & 222 & 1 & 13 24 46.0 & --47 30 14 & 16.68 & 1.07 \\
 5\_4\_1514 & 20.4 & 209 & 1 & 13 27 09.5 & --47 08 34 & 16.47 & 0.93 \\
 \enddata
 \tablecomments{Table \ref{members} is published in its entirety in the electronic addition of the
 Astronomical Journal.  A portion is shown here for guidance regarding its form and content.}
 \end{deluxetable}

\begin{deluxetable}{rccccccc}
\tablewidth{0pt}
\tablecaption{Probable non-cluster-member stars \label{nonmembers}}
\tablecolumns{8}
\tablehead{
\colhead{ID} & \colhead{R} & \colhead{Velocity} & \colhead{N}  & \colhead{RA} & \colhead{Dec} & 
\colhead{$V$} & \colhead{$V-I$} \\
 & \colhead{(arcmin)} & \colhead{(km s$^{-1}$)} & & \colhead{(J2000)} & \colhead{(J2000)}  &  & }
 \startdata
 7\_2\_312 & 20.7 & 197 & 2 & 13 24 44.4 & --47 25 56 & 16.21 & 0.96  \\
 11\_1\_4329 & 21.4 & 226 & 1 & 13 27 27.7 & --47 48 47 & 16.53 &  0.96 \\
 8\_1\_1381 & 22.5 & 232 & 1 & 13 28 24.3 & --47 13 33 & 16.55 & 0.88 \\
 11\_8\_1684 & 23.0 & 232 & 1 & 13 26 00.2 & --47 50 14 & 16.06 & 1.04 \\
 11\_1\_3775 & 32.5 & 219 & 1 & 13 27 39.0 & --47 59 55 & 15.56 & 1.22 \\
 5\_6\_92 & 34.9 & 202 & 1 & 13 26 36.1 & --46 53 48 & 16.06 & 1.03 \\
 12\_8\_1988 & 37.9 & 206 & 1 & 13 29 30.5 & --47 54 25 & 16.35 & 1.15 \\
 6\_6\_1259 & 45.0 & 197 & 2 & 13 29 56.3 & --46 57 16 & 16.73 & 0.89 \\
 10\_2\_28 & 47.3 & 260 & 1 & 13 24 45.8 & --48 11 24 & 16.17 & 1.14 \\
 10\_1\_2777 & 48.0 & 254 & 1 & 13 22 31.9 & --47 50 17 & 16.43 & 0.88 \\
 4\_3\_2022 & 54.6 & 256 & 2 & 13 23 07.1 & --46 48 42 & 16.15 & 1.21 \\
 11\_3\_191 & 55.1 & 243 & 1 & 13 29 17.7 & --48 17 31 & 15.54 & 1.16 \\
 \enddata
 \tablecomments{Star 6\_6\_1259 excluded on basis of discrepant proper motion; all others
 excluded on the basis of discrepant line strengths; star 10\_2\_28 is a carbon star.}
 \end{deluxetable}
 
 \begin{deluxetable}{rcrrcccc}
\tablewidth{0pt}
\tablecaption{Non-member stars \label{vel_nonmems}}
\tablecolumns{8}
\tablehead{
\colhead{ID} & \colhead{R} & \colhead{Velocity} & \colhead{N}  & \colhead{RA} & \colhead{Dec} & 
\colhead{$V$} & \colhead{$V-I$} \\
 & \colhead{(arcmin)} & \colhead{(km s$^{-1}$)} & & \colhead{(J2000)} & \colhead{(J2000)}  &  & }
 \startdata
11\_8\_135 & 20.0 & --11 & 2 & 13 26 44.3 & --47 48 38 & 16.05 & 1.02  \\
 8\_4\_4179 & 20.2 & --42 & 1 & 13 27 31.3 & --47 47 16 & 16.43 &  1.02 \\
11\_1\_5863 & 20.2 & 34 & 10 & 13 26 59.0 & --47 48 44 & 16.23 & 1.08 \\
 8\_8\_1220 & 20.5 & --5 & 8 & 13 26 21.6 & --47 08 30 & 16.39 & 0.85 \\
 8\_1\_1736 & 20.6 & --11 & 1 & 13 28 14.9 & --47 14 39 & 16.24 & 0.94 \\
 8\_2\_2607 & 20.6 & 168 & 1 & 13 28 23.8 & --47 16 18 & 15.67 & 1.05 \\
 8\_8\_2721 & 20.6 & 26 & 1 & 13 25 55.6 & --47 09 52 & 16.65 & 0.95 \\
 8\_5\_4726 & 20.7 & 6 & 1 & 13 25 17.3 & --47 43 02 & 16.29 & 1.03 \\
 8\_8\_4580 & 20.7 & --19 & 1 & 13 25 12.7 & --47 15 09 & 16.64 & 0.97 \\
 5\_4\_1370 & 20.9 & 55 & 8 & 13 27 19.0 & --47 08 30 & 16.19 & 0.99 \\
 \enddata
 \tablecomments{Table \ref{vel_nonmems} is published in its entirety in the electronic addition of the
 Astronomical Journal.  A portion is shown here for guidance regarding its form and content.}
 \end{deluxetable}
 
 \begin{deluxetable}{rccccccc}
\tablewidth{0pt}
\tablecaption{Candidate extra-tidal cluster members \label{tab_4}}
\tablecolumns{8}
\tablehead{
\colhead{ID} & \colhead{R} & \colhead{Velocity} & \colhead{N}  & \colhead{RA} & \colhead{Dec} & 
\colhead{$V$} & \colhead{$V-I$} \\
 & \colhead{(arcmin)} & \colhead{(km s$^{-1}$)} & & \colhead{(J2000)} & \colhead{(J2000)}  &  & }
 \startdata
 4\_7\_288 & 66.5 & 238 & 1 & 13 22 03.5 & --46 42 37 & 16.43 & 0.98  \\
 7\_8\_1878 & 70.1 & 238 & 1 & 13 20 07.1 & --47 09 50 & 15.70 &  1.28 \\
 14\_3\_2302 & 93.0 & 246 & 1 & 13 27 49.7 & --49 01 00 & 16.57 & 1.08 \\
 1\_2\_745 & 99.0 & 230 & 1 & 13 23 42.4 & --45 54 46 & 15.90 & 1.18 \\
 15\_5\_941 & 112.6 & 219 & 1 & 13 30 36.3 & --49 14 28 & 16.50 & 1.02 \\
 3\_8\_141 & 113.3 & 227 & 1 & 13 31 01.3 & --45 44 08 & 16.42 & 1.01 \\
 \enddata
 \end{deluxetable}
 
 \begin{deluxetable}{cccc}
 \tablewidth{0pt}
 \tablecaption{$\omega$ Cen surface density data \label{sb_tab}}
 \tablecolumns{4}
 \tablehead{
 \colhead{log r} & \colhead{log SD} & \colhead{log r} & \colhead{log SD}\\
 \colhead{[arcmin]} & \colhead{[10.0 $V$ mag/arcmin$^2$]} & \colhead{[arcmin]} & \colhead{[10.0 $V$ mag/arcmin$^2$]}}
 \startdata
 1.115 & --2.076 $\pm$ 0.011& 1.323 & --2.769  $\pm$ 0.063 \\
 1.177 & --2.308 $\pm$ 0.014 & 1.362 & --3.022 $\pm$ 0.080 \\
 1.231 & --2.449 $\pm$ 0.016 & 1.398 & --3.165 $\pm$ 0.090\\
 1.279 & --2.674 $\pm$ 0.022 & 1.448 & --3.806  $\pm$ 0.129\\
 1.323 & --2.809 $\pm$ 0.026 & 1.513 &  --3.871 $\pm$ 0.115\\
 1.362 & --2.928 $\pm$ 0.031 & 1.575 & --3.933 $\pm$ 0.115\\
 1.398 & --3.090 $\pm$ 0.040  & 1.629 & --4.561$\pm$ 0.239\\
 1.432 & --3.164 $\pm$ 0.044 & 1.677 & --4.734 $\pm$ 0.286\\
 1.463 & --3.343 $\pm$ 0.061 & 1.729 &  --5.108 $\pm$ 0.383 \\
 1.492 & --3.422 $\pm$ 0.069 & \nodata & \nodata \\
 1.532 & --3.766 $\pm$ 0.097 & \nodata & \nodata \\
 1.589 & --4.019 $\pm$ 0.160 & \nodata & \nodata \\
 \enddata
 \tablecomments{The first 2 columns are from \citet{MC07}, the remainder from the 
 cluster member sample determined here.  The surface densities have the same normalization
 as Table 1 of \citet{GM87}.  Add 0.832 (equivalent to a central surface brightness
 of 16.81 {\it V} mag/arcsec$^2$) for log(surface density) values in the adopted units.}
 \end{deluxetable}

\end{document}